\documentclass[aps,prd,twocolumn,reprint,preprintnumbers,nofootinbib]{revtex4}

\usepackage{amsmath}
\usepackage{amssymb}
\usepackage{bbm}
\usepackage{feynmp}
\usepackage{graphicx}
\usepackage{hyperref}
\usepackage{siunitx}
\usepackage{slashed}
\usepackage{subfigure}
\usepackage[usenames,dvipsnames,svgnames]{xcolor}

\newcommand{\krf}{|\vec{k}_{\rm RF}|}

\renewcommand{\Im}[1]{{\rm Im}\left\lbrace{#1}\right\rbrace}
\renewcommand{\Re}[1]{{\rm Re}\left\lbrace{#1}\right\rbrace}
\DeclareMathOperator{\tr}{Tr}
\DeclareMathOperator{\Res}{Res}

\newcommand{\dd}{{\rm d}}

\newcommand{\refapp}[1]{app.~\ref{app:#1}}
\newcommand{\refeq}[1]{eq.~(\ref{eq:#1})}
\newcommand{\refeqs}[2]{eqs.~(\ref{eq:#1})-(\ref{eq:#2})}
\newcommand{\reffig}[1]{fig.~\ref{fig:#1}}
\newcommand{\refsec}[1]{section~\ref{sec:#1}}

\newcommand{\eps}{\varepsilon}
\newcommand{\para}{\parallel}
\newcommand{\calM}{\mathcal{M}}
\newcommand{\order}[1]{\mathcal{O}\left({#1}\right)}

\newcommand{\ie}{\textit{i.e.}}

\numberwithin{equation}{section}
\allowdisplaybreaks

\begin{document}

\title{Disentangling the Decay Observables in $B^-\to \pi^+\pi^-\ell^-\bar\nu_\ell$}
\author{Sven Faller}
\email{sven.faller@uni-siegen.de}
\author{Thorsten Feldmann}
\email{thorsten.feldmann@uni-siegen.de}
\author{Alexander Khodjamirian}
\email{khodjamirian@physik.uni-siegen.de}
\author{Thomas Mannel}
\email{mannel@physik.uni-siegen.de}
\author{Danny van Dyk}
\email{vandyk@physik.uni-siegen.de}
\affiliation{Theoretische Physik 1, Naturwissenschaftlich-Technische Fakult\"at,
Universit\"at Siegen, Walter-Flex-Stra\ss{}e 3, D-57068 Siegen, Germany}
\date{\today}

\preprint{SI-HEP-2013-07, QFET-2013-06}

\begin{abstract}
We study the semileptonic $b\to u$ transition in the decay mode $B^-\to \pi^+\pi^-\ell^-\bar\nu_\ell$.
We define $B\to \pi \pi$ form factors in the helicity basis, and study their properties in
various kinematic limits, including form factor relations in the heavy-mass and large-energy limits,
the decomposition into partial waves of the dipion system, and the resonant contribution of
vector and scalar mesons. We show how angular observables in $B^-\to \pi^+\pi^-\ell^-\bar\nu_\ell$
can be used to measure dipion form factors or to perform null tests of the Standard Model.
\end{abstract}

\maketitle

\section{Introduction}

The decay $B \to \pi\pi \ell^-\bar\nu_\ell$ is interesting for several reasons.
At quark level, it is generated by the semi-leptonic $b \to u\ell\nu_\ell$ transition which,
in the Standard Model (SM), is induced by tree-level $W$-boson exchange but proportional
to the small element $V_{ub}$ of the Cabibbo-Kobayashi-Maskawa (CKM) matrix.
For a while the common paradigm has been to search for new physics (NP) in rare
\emph{loop-induced} flavour transitions. Meanwhile, in light of the small tension observed between
the determinations of $|V_{ub}|$ from inclusive $B \to X_u\ell^-\bar\nu_\ell$ or
exclusive semi-leptonic $B\to \lbrace \pi,\rho\rbrace \ell^-\bar\nu_\ell$ decays
\cite{Kowalewski:2012:PDG,Amhis:2012bh}, systematic tests of $b \to u$ transitions in the SM and beyond
appear timely. In this context,
the dipion system in the hadronic final state not only provides an independent decay
channel, but, more importantly, offers the possibility to explore a number of angular
observables that are sensitive to the spin structure of the underlying short-distance
operators responsible for the decay in the SM or NP. The situation here
is similar to the analysis 
of rare $b \to s$ transitions in $B \to (K\pi)_{S,P}\ell^+\ell^-$ decays, see for example
\cite{Kruger:2005ep,Altmannshofer:2008dz,Bobeth:2010wg,Lu:2011jm,Becirevic:2012dp,Matias:2012qz,Blake:2012mb,Bobeth:2012vn,Doring:2013wka}.

Moreover, the phase space associated with the 
kinematics of the four-body decay covers various limiting cases for which 
specific theoretical approaches to handle the strong-interaction effects in
Quantum Chromodynamics (QCD) are applicable.
In particular, this includes expansions in small light-quark or large heavy-quark
masses based on effective-field theory methods. 
For instance, the case of two pions recoiling
against each other with a large energy can be used 
to assess the reliability of theoretical predictions in the 
QCD factorization approach which has been frequently used for
non-leptonic $B \to \pi\pi$ decays \cite{Beneke:1999br,Beneke:2001ev}.
The decay $B^- \to \pi^+\pi^- \ell^- \bar\nu_\ell$ also involves the resonant channel
$B^- \to \rho^0(\to\pi^+\pi^-)\ell^-\bar\nu_\ell$ decay which is one of the 
aforementioned exclusive modes where the $|V_{ub}|$-extraction is not in
perfect agreement with the inclusive determination. The theoretical exploration of the
various corners of (non-resonant) phase space will therefore also help to better 
understand the proper description of the $B \to \rho\ell\bar\nu_\ell$ decay beyond the
approximation of narrow width and flat non-resonant background.

Our paper is organized as follows.
In the following \refsec{decay}, we provide the basic definitions for
$B \to \pi\pi$ form factors that are most convenient for the
angular analysis  and for the theoretical description of the decay
in certain kinematic limits.
In \refsec{limits} we consider the dipion form factors
for two kinematic limits,
giving rise to symmetry relations in heavy-quark effective theory (HQET),
and soft-collinear effective theory (SCET), respectively.
Further form factor properties in specific kinematic situations,
namely the perturbative factorization in the limit of almost back-to-back
energetic pions, on the one hand, and the description of hadronic resonances
in the dipion channel, on the other hand, are the subject of
\refsec{props}.
The phenomenology of the angular distributions 
of the decay $B^-\to\pi^+\pi^-\ell^-\bar\nu_\ell$ in the SM 
is worked out in \refsec{decay:obs}. In \refsec{model-indep-results}
we combine knowledge of the form
factor limits with the angular distribution to derive
relations between the angular observables that
do not depend on a hadronic model. We conclude in \refsec{conclusion}.

\section{$B\to \pi\pi$ Form Factors}
\label{sec:decay}

\subsection{Kinematics}

Let us begin with the definition of the
kinematics. In the following, 
$p^\mu=M_B v^\mu$ will denote the 4-momentum
of the decaying $B$-meson. The projection with its four-velocity $v^\mu$
defines the energy of the final-state particles
in the $B$-meson rest frame,
$p^0 = (v \cdot p)$.
The momenta of the decay products will be denoted as $k_1^\mu,k_2^\mu$
for the two pions, and $q_1^\mu,q_2^\mu$ for the two leptons,
with the specific charge assignment according to
\begin{align*}
    B^-(p) &\to \pi^+(k_1) \, \pi^-(k_2) \, \bar\nu(q_1) \, \ell^-(q_2) \,.
    \label{eq:hadkin}
\end{align*}
We define the sum and difference of hadronic and leptonic momenta as
\begin{equation}
\begin{aligned}
    q                       & = q_1 + q_2\,, \qquad\qquad
    k                         = k_1 + k_2\,,  \cr 
    \bar{q}                 & = q_1 - q_2\,, \qquad\qquad
    \bar{k}                   = k_1 - k_2\,.
\end{aligned}
\end{equation}
The hadronic system is then described by three kinematic Lorentz invariants:
the momentum transfer $q^2$, the dipion invariant mass $k^2$,
and the scalar product $q \cdot \bar{k}$. The latter defines the polar angle $\theta_\pi$
of the $\pi^+$ in the dipion rest frame
\begin{align}
 q \cdot \bar{k}  &= \frac{\beta_\pi}{2} \, \sqrt{\lambda} \, \cos\theta_\pi\,,
\end{align}
where $\beta_\pi^2 = (k^2-4M_\pi^2)/k^2 = -\bar{k}^2/k^2$,
and $\lambda = \lambda(M_B^2, q^2, k^2)$ is the K\"all\'en function
\begin{equation}
    \lambda(a, b, c) = a^2 + b^2 + c^2 - 2(a b + b c + c a)\,.
    \label{eq:kaellen-fkt}
\end{equation}
The relative orientation between the leptons
and hadrons in the final state is further characterized by the Lorentz invariants
\begin{equation}
\begin{aligned}
 k\cdot \bar q &= \frac12 \, \sqrt{\lambda} \, \cos\theta_\ell \,,
 \cr 
 \bar k \cdot \bar q &= \frac{\beta_\pi}{2} \Big( (M_B^2 - k^2 - q^2) \,
\cos\theta_\ell \, \cos \theta_\pi \cr & \qquad \quad  - 2 \, \sqrt{q^2 k^2} \, \sin\theta_\ell \,
\sin\theta_\pi \, \cos\phi \Big)
 \,,
\end{aligned}
\end{equation}
where $\theta_\ell$ is the polar angle of the negatively charged lepton in
the dilepton rest frame, and $\phi$ is the azimuthal angle between the 
dilepton and dipion decay plane.
Here and in the following lepton masses are set to zero.
More details can be found in \refapp{kin}.

In the following, it will be convenient to construct an orthogonal basis of 
momentum vectors, 
\begin{equation}
\begin{aligned}
    q^\mu & \,,\cr 
    k_{(0)}^\mu             & = k^\mu - \frac{k \cdot q}{q^2} q^\mu\,, \cr 
    \bar{k}_{(\para)}^\mu   & = \bar{k}^\mu - \frac{4 (k\cdot q) (q \cdot \bar{k})}{\lambda} \, k^\mu
                             + \frac{4 k^2 (q\cdot \bar{k})}{\lambda} \, q^\mu\,, \cr 
    \bar{q}_{(\perp)}^\mu   &= 2 \, \epsilon^{\mu\alpha\beta\gamma} \, 
    \frac{q_\alpha \, k_\beta \, \bar k_\gamma}{\sqrt{\lambda}}\,.
\end{aligned}
\label{eq:ortho}
\end{equation}
Properly normalized, using
\begin{align}
& k_{(0)}^2 = - \frac{\lambda}{4q^2} \,, \qquad 
\bar k_{(\para)}^2 = \bar q_{(\perp)}^2 = - \beta_\pi^2 \, k^2 \sin^2\theta_\pi \,,
\end{align}
the vectors in \refeq{ortho} represent an orthonormal basis of time-like and space-like
polarization vectors associated with the 
leptonic currents, see also \refeq{polvec-dilepton-rf} in the appendix,
\begin{equation}
\begin{aligned}
 \eps^\mu(t) &= \frac{1}{\sqrt{q^2}} \, q^\mu \,,
 \qquad 
 \eps^\mu(0) =-\frac{2 \sqrt{q^2}}{\sqrt\lambda} \, k_{(0)}^\mu \,,
 \cr
 \eps^\mu(\pm) &= - \frac{1}{\sqrt{2k^2} \, \beta_\pi \, \sin\theta_\pi} 
 \left( \bar k_{(\para)}^\mu
  \mp i \, \bar q_{(\perp)}^\mu \right)\!e^{\mp i\phi} \,,
\end{aligned}
  \label{eq:epsdefcov}
\end{equation}
which will be used to project onto helicity form factors.

\subsection{Vector and  Axial-Vector Form Factors}

In the SM, the $B \to \pi\pi\ell\nu$ decay amplitudes are characterized by the transition form factors
for vector and axial-vector $b\to u$ currents between a $B$-meson and two pions.
Using the definitions of the previous subsection, 
we parametrize the hadronic matrix elements 
in terms of one vector form factor $F_\perp$, 
\begin{equation}
    \langle \pi^+(k_1) \pi^-(k_2)|\bar{u}\gamma^\mu b|B^-(p)\rangle
    = i F_\perp \, \frac{1}{\sqrt{k^2}} \ \bar q_{(\perp)}^\mu
    \,,
    \label{eq:decay:hme:param1}
\end{equation}
and three axial-vector form factors
$F_t$, $F_0$, $F_\para$,
\begin{multline}
    -\langle \pi^+(k_1) \pi^-(k_2)|\bar{u}\gamma^\mu \gamma_5 b|B^-(p)\rangle\\
    = F_t \, \frac{q^\mu}{\sqrt{q^2}} 
    + F_0 \, \frac{2\sqrt{q^2}}{\sqrt{\lambda}} \, k_{(0)}^\mu
    + F_\para \, \frac{1}{\sqrt{k^2}} \, \bar{k}_{(\para)}^\mu\,.
\label{eq:decay:hme:param2}
\end{multline}
Note here, that the apparent divergence of the hadronic matrix elements in the limit
$q^2\to 0$ is compensated by an appropriate phase space factor, see \refeq{def-trans-amp}
and \refeq{def-phase-space-N}.
Here, each form factor depends on the three independent Lorentz invariants $q^2$, $k^2$ and $q\cdot \bar{k}$.
It is also to be noted
that, in general, the dipion form factors are \emph{complex} functions 
above threshold $k^2> 4m_\pi^2$.
The prefactors in \refeq{decay:hme:param1} and \refeq{decay:hme:param2}
are chosen in such a way that the form factors correspond to particular
helicity amplitudes which can be simply obtained by 
contraction
\begin{align}
H_{\lambda} &\equiv \langle \pi^+ \pi^-|\bar{u} \gamma^\mu (1 - \gamma_5) b|B^-\rangle 
\, \eps_\mu^\dagger(\lambda) \,,
\end{align} 
with  the polarization vectors 
as defined in \refeq{epsdefcov}.
We obtain
\begin{equation}
\begin{aligned}
    H_t           & = F_t\,,\qquad\qquad H_0 = F_0\,,\cr 
    H_{\pm}       & = (F_\para \pm F_\perp  ) \,
      \frac{\beta_\pi}{\sqrt{2}} \, \sin\theta_\pi \, e^{\pm i\phi}\,.
\end{aligned}
    \label{eq:decay:hme:helicity-amps}
\end{equation}
In terms of the so-defined ``helicity form factors'', one obtains simple expressions for 
the differential decay widths in the angular analysis and simple relations between form factors 
in HQET or SCET, which have also been emphasized for other decay modes
\cite{Bharucha:2010im,Bobeth:2010wg,Feldmann:2011xf,Das:2012kz,Hambrock:2012dg}.

\subsection{Partial Waves}

The $B\to\pi\pi$ helicity amplitudes can be expanded in terms of associated Legendre polynomials
$P_\ell^{(m)}(\cos\theta_\pi)$, with $\ell=(0,1,2,\ldots)$ corresponding to $(S,P,D,\ldots)$ partial waves.
For the helicity amplitudes $H_0$ and $H_t$ one obtains
\begin{equation}
\begin{aligned}
    H_{0,t} 
        & = \sum\limits_{\ell=0}^\infty \sqrt{2\ell\!+\!1} \, H_{0,t}^{(\ell)}(q^2, k^2) \, P_\ell^{(0)}(\cos\theta_\pi)\\
        & = H_{0,t}^{(S)}(q^2, k^2) + \sqrt{3} \, H_{0,t}^{(P)}(q^2, k^2) \, \cos\theta_\pi\\
        & + \ldots\,,
\end{aligned}
    \label{eq:Hexp1}
\end{equation}
and for $H_\pm$ one gets
\begin{align}
    H_{\pm} \notag
        & = \sum\limits_{\ell=1}^\infty \sqrt{2\ell\!+\!1} \, H_\pm^{(\ell)}(q^2, k^2) \, P_\ell^{(\pm 1)}(\cos\theta_\pi)
         \, e^{\pm i\phi}\\
        & = \mp \frac{\sqrt3}{\sqrt 2} \,  H_\pm^{(P)}(q^2, k^2) \, \sin\theta_\pi \, e^{\pm i\phi} + \ldots\,, \label{eq:Hexp2}
\end{align}
which contains no $S$-wave contribution.
For the form factors $F_0$ and $F_t$ this directly translates into 
the partial-wave expansion
\begin{equation}
    \label{eq:decay:hme:ff-legendre-longitudinal}
\begin{aligned}
    F_{0,t}
        & = \sum\limits_{\ell=0}^\infty \sqrt{2\ell + 1} \, F_{0,t}^{(\ell)}(q^2, k^2) \, P_\ell^{(0)}(\cos\theta_\pi)\cr 
        & = F_{0,t}^{(S)}(q^2, k^2) + \sqrt{3} \, F_{0,t}^{(P)}(q^2, k^2) \, \cos\theta_\pi\\
        & + \ldots
\end{aligned}
\end{equation}
so that $H_{0,t}^{(\ell)} = F_{0,t}^{(\ell)}$.
For the form factors $F_\para$ and $F_\perp$ we define
 \begin{align}
   F_{\para,\perp}
    & = - \sum\limits_{\ell=1}^\infty \sqrt{2\ell + 1} \, F_{\para,\perp}^{(\ell)}(q^2, k^2)
      \, \frac{P_\ell^{(1)}(\cos\theta_\pi)}{\sin\theta_\pi} \cr 
        & = \frac{\sqrt{3}}{\sqrt{2}} \, F_{\para,\perp}^{(P)}(q^2, k^2) 
          + \ldots
          \label{eq:decay:hme:ff-legendre-transverse}
\end{align}
such that $H_\pm^{(\ell)} = \mp \frac{\beta_\pi}{\sqrt2} \left( F_\para^{(\ell)} \pm F_\perp^{(\ell)} \right)$.

\section{Form Factor Relations}
\label{sec:limits}

In certain kinematic limits, the form factors will obey 
approximate symmetry relations which would become exact
in the limit of infinitely heavy $b$-quark mass.
Similar to what is known from $B \to K^{(*)}\ell^+\ell^-$ decays,
the form factor relations allow relatively robust predictions for 
angular observables which are independent of hadronic matrix elements
in these limits.
Note that the form factor relations are valid for each
partial wave separately.

\subsection{HQET Limit}
\label{sec:limits:hqet-chipt}

\begin{figure}
    \includegraphics[width=.49\textwidth]{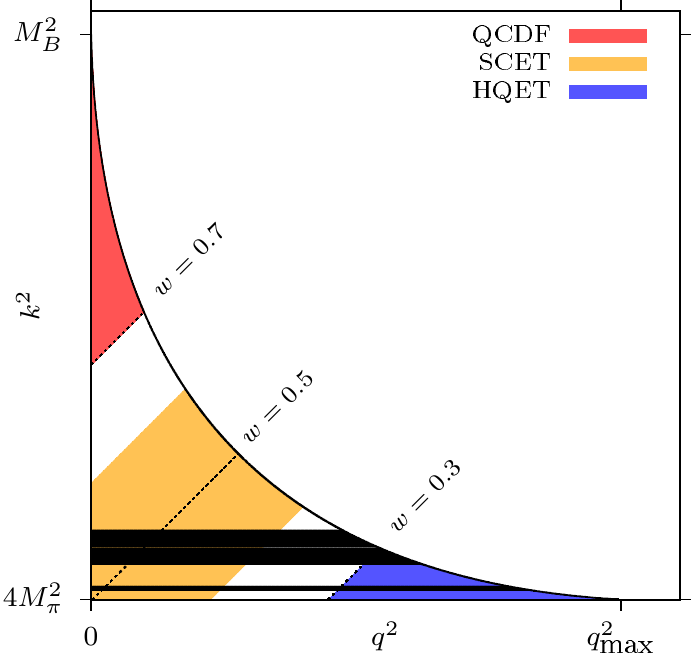}
    \caption[]{(color online) Sketch of the $q^2$-$k^2$ phase space
        with $w = (v\cdot k) / M_B$ isolines for $w = 0.3,\,0.5,\,0.7$ (dashed lines).
        The typical regions of applicability for the different theory approaches,
        labelled QCDF ($w \geq 0.7$, red),
        SCET ($0.6 \geq w \geq 0.4$, orange) and HQET ($0.3 \geq w$, blue), are highlighted.
        The $\rho$ resonances $\rho(770)$, $\rho(1450)$, $\rho(1700)$ are overlaid as horizontal grey bands
        for illustrative purpose.
        \label{fig:phase-space}
    }
\end{figure}

If the energy-transfer to the hadronic final state is small, i.e.
$(v \cdot k) \sim \Lambda_{\rm had} \ll m_b$, the heavy $b$-quark acts as a quasi-static
source of color, and the techniques of HQET are applicable.
For the kinematic invariants in the $\pi\pi$-system this implies,
\begin{align}
    q^2 & \sim m_b^2\,, & k^2 & \sim \Lambda_{\rm had}^2 \,,      & (q\cdot k) \sim \Lambda_{\rm had} m_b \,.
\end{align}
with $\Lambda_\text{had}$ being a typical hadronic scale of order of a few hundred $\SI{}{\mega\electronvolt}$;
see also \reffig{phase-space} for a sketch of the phase space.
In particular, the general set of heavy-to-light form factors for arbitrary Dirac structures can
be related to a smaller set of Isgur-Wise functions \cite{Isgur:1990kf,Falk:1990yz}. To this end, the dipion system is
represented by the most general Dirac structure that can be constructed from the two pion momenta
and the heavy-quark velocity.
We define the following parameterization, 
\begin{align}
    {\cal M}_{\pi\pi}(k,\bar k, v) & \equiv
      \Xi_1 \, \frac{1}{\sqrt{q^2}} \, \slashed q
    + \Xi_2 \, \frac{2\sqrt{q^2} }{\sqrt{\lambda}}\, \slashed{k}_{(0)}\\
    & \quad 
    + \Xi_3 \, \frac{1}{\sqrt{k^2}} \, \bar{\slashed{k}}_{(\para)}
    + i \, \Xi_4 \, \frac{1}{\sqrt{k^2}}\, \bar{\slashed{q}}_{(\perp)} \! \gamma_5\,,\notag
\end{align}
which introduces four independent Isgur-Wise functions $\Xi_i = \Xi_i(v\cdot k, k^2, \cos\theta_\pi)$.
For a given decay current, the form factors can then be obtained in terms of Clebsch-Gordan coefficients 
given by the Dirac trace,
\begin{multline}
    \langle \pi^+(k_1)\pi^-(k_2)| \bar u \, \Gamma \, h_v^{(b)}|B^-(p)\rangle\\
        = \frac12 \, \tr \left[ {\cal M}_{\pi\pi} \, \Gamma \, \frac{1+\slashed v}{2} \, (-\gamma_5) \right] \,,
    \label{eq:HQETtrace}
\end{multline}
where $\Gamma$ is the Dirac matrix of the underlying current.
For left-handed SM currents this yields one-to-one relations
between the four helicity form factors $F_i$ and the Isgur-Wise functions,
\begin{align}
 F_t &= \Xi_1 \,, \quad F_0 = \Xi_2 \,, \quad 
 F_\parallel = \Xi_3 \,, \quad F_\perp = \Xi_4 \,.
\end{align}
In the presence of NP, other $b \to u \ell\nu$ operators may contribute, and the
corresponding form factors for pseudoscalar, tensor or pseudotensor currents would be given by
the same set of Isgur-Wise functions $\Xi_{1-4}$.

Explicit theoretical expressions 
for the Isgur-Wise functions $\Xi_{1-4}$ can be obtained in the limit
where the two pions are soft, $v \cdot k_i \sim M_\pi$, 
in which case the methods of heavy-meson
chiral perturbation theory \cite{Burdman:1992gh} are applicable.

\subsection{SCET Limit}
\label{sec:limits:scet-dd}

If the energy-transfer to the hadronic final state is large, $(v\cdot k)\sim m_b/2 \gg \Lambda_{\rm had}$,
while the invariant mass is small, $k^2 \ll m_b^2$,
which also implies $q^2 \ll m_b^2$, the hadronic dynamics can be treated
in soft-collinear effective theory (SCET) \cite{Bauer:2000yr,Beneke:2002ph}.
The phase space region associated with this limit is sketched in \reffig{phase-space}.
Similarly to the HQET case, this yields new form-factor
symmetry relations which have already been established for single light pseudoscalars
or vector mesons in the final state \cite{Charles:1998dr,Beneke:2000wa} (analogous relations
for baryonic decays can be found in \cite{Mannel:2011xg,Feldmann:2011xf}). 
These can be conveniently derived by introducing light-like vectors 
$n_\pm^\mu$ in the $k$--$q$ plane, according to
\begin{equation}
    n_\pm^\mu = \left(1 \mp \frac{1}{\eta}\right) v^\mu \pm \frac{1}{|\vec{q}\,|}q^\mu\,,
\end{equation}
in terms of 
the rapidity $\eta$ and the three-momentum $|\vec{q}\,|$ of the lepton pair in the $B$-meson rest frame:
\begin{equation}
    \eta       = \frac{\sqrt{\lambda}}{M_B^2 - k^2 + q^2}\,,\qquad 
    |\vec{q}\,|  = \frac{\sqrt{\lambda}}{2 M_B}\,.
\end{equation}
These vectors satisfy the relations
\begin{equation}
        n_\pm^2  = 0\,,  \qquad    n_+ \cdot n_-  = 2\,, \qquad  n_+^\mu + n_-^\mu = 2 v^\mu\,,
\end{equation}
and can be used to construct Dirac projectors
\begin{equation}
 P_\pm = \frac{\slashed n_\pm \slashed n_\mp}{4} \,.
\end{equation}
In the large-energy limit, only the $P_+$ projection of the
energetic $u$-quark in the $b\to u \ell \nu$ transition contributes.
The trace in \refeq{HQETtrace} then simplifies further, because 
the terms with $\slashed q$ and $\slashed k_{(0)}$
($\bar{\slashed k}_{(\para)}$ and $\bar{\slashed q}_{(\perp)}$) in ${\cal M}_{\pi\pi}$
yield the same contribution,
\begin{multline}
    \langle \pi^+(k_1)\pi^-(k_2)| \bar u \, \Gamma \, h_v^{(b)}|B^-(p)\rangle\\
        = \tr\left[\left(
            \frac{\slashed q}{\sqrt{q^2}} \, \xi_L 
            + \frac{\bar{\slashed{k}}_{(\para)}}{\sqrt{k^2}} \, \xi_T \right) 
            P_+ \, \Gamma \, \frac{1 + \slashed{v}}{2} \, (-\gamma_5)\right]\,,
    \label{eq:SCETtrace}
\end{multline}
which implies the large-recoil form-factor relations
\begin{equation}
\begin{aligned}
 F_t = F_0 = \Xi_1=\Xi_2 &\equiv \xi_L \,,\\
 F_\para= F_\perp = \Xi_3=\Xi_4 & \equiv \xi_T \,.
\end{aligned}
\end{equation}

Theoretical approaches to predict the form factors $\xi_L$ and $\xi_T$ in the SCET limit depend
on the distribution of the large energy/momentum among the two pions:
\begin{itemize}
 \item If both pions are energetic and move collinear with a small invariant mass $k^2 \sim \Lambda_{\rm had}^2$,
   the 2-pion state could be described by a generalized distribution amplitudes (GDAs),
   i.e.\ a two-pion light cone distribution amplitudes (2$\pi$LCDAs)
   \cite{Mueller:1998fv,Diehl:1998dk,Polyakov:1998ze,Diehl:2000uv}.
   The 2$\pi$LCDAs contain the time-like pion form factors and the contributing
   hadronic resonances (notably $\rho \to \pi\pi$) as a limiting case.
   
 \item If only one pion is energetic and the other soft, a combination of
    SCET/QCDF and chiral perturbation theory should apply, similar to
    \cite{Grinstein:2005ud} where this combination was studied in the context
    of $\bar{B}\to\bar{K}\pi\ell^+\ell^-$ decays.
 
\end{itemize}

\section{Form Factor Properties}
\label{sec:props}

In this section we briefly comment on further generic properties of
the dipion form factors that are characteristic in certain regions
of the $|\pi\pi\rangle$ phase space.

\subsection{QCD Factorization for Large Dipion Masses}
\label{sec:props:QCDF}

\begin{figure*}[t]
\begin{center}
    \begin{fmffile}{feyn-btouqbarqlnu-qcdf}
    \subfigure[]{
    \resizebox{.25\textwidth}{!}{
    \begin{fmfgraph*}(200,100)
        \fmfforce{(0,.70h)}{ib}
        \fmfforce{(0,.10h)}{is}
        \fmfforce{(.33w,h)}{ow}
        \fmfforce{(.33w,.70h)}{veff}
        \fmfforce{(w,.70h)}{ou}
        \fmfforce{(w,.50h)}{oqbar}
        \fmfforce{(w,.30h)}{oq}
        \fmfforce{(w,.10h)}{os}
        \fmfforce{(.8w,.40h)}{vg2}
        \fmf{fermion}{ib,veff,vg1,ou}
        \fmffreeze
        \fmf{fermion}{oqbar,vg2,oq}
        \fmf{phantom}{veff,vg1}
        \fmf{fermion,tension=0}{os,is}
        \fmf{dashes}{veff,ow}
        \fmf{gluon}{vg1,vg2}
        \fmfv{d.shape=square,d.size=.05h}{veff}
        \fmfv{label=$b$}{ib}
        \fmfv{label=$\bar{u}$}{is}
        \fmfv{label=$u$}{ou}
        \fmfv{label=$\bar{d}$}{oqbar}
        \fmfv{label=$d$}{oq}
        \fmfv{label=$\bar{u}$}{os}
        \fmfv{label=$q$,l.angle=0}{ow}
    \end{fmfgraph*}}}
    \qquad
    \subfigure[]{
    \resizebox{.25\textwidth}{!}{
    \begin{fmfgraph*}(200,100)
        \fmfforce{(0,.70h)}{ib}
        \fmfforce{(0,.10h)}{is}
        \fmfforce{(.33w,.70h)}{vg1}
        \fmfforce{(.66w,h)}{ow}
        \fmfforce{(.66w,.70h)}{veff}
        \fmfforce{(w,.70h)}{ou}
        \fmfforce{(w,.50h)}{oqbar}
        \fmfforce{(w,.30h)}{oq}
        \fmfforce{(w,.10h)}{os}
        \fmfforce{(.8w,.40h)}{vg2}
        \fmf{fermion}{ib,vg1,veff,ou}
        \fmffreeze
        \fmf{fermion}{oqbar,vg2,oq}
        \fmf{phantom}{veff,vg1}
        \fmf{fermion,tension=0}{os,is}
        \fmf{dashes}{veff,ow}
        \fmf{gluon,right=.3}{vg1,vg2}
        \fmfv{d.shape=square,d.size=.05h}{veff}
        \fmfv{label=$b$}{ib}
        \fmfv{label=$\bar{u}$}{is}
        \fmfv{label=$u$}{ou}
        \fmfv{label=$\bar{d}$}{oqbar}
        \fmfv{label=$d$}{oq}
        \fmfv{label=$\bar{u}$}{os}
        \fmfv{label=$q$,l.angle=0}{ow}
    \end{fmfgraph*}}}
    \qquad
    \subfigure[]{
    \resizebox{.25\textwidth}{!}{
    \begin{fmfgraph*}(200,100)
        \fmfforce{(0,.70h)}{ib}
        \fmfforce{(0,.10h)}{is}
        \fmfforce{(.33w,h)}{ow}
        \fmfforce{(.33w,.70h)}{veff}
        \fmfforce{(w,.70h)}{ou}
        \fmfforce{(w,.50h)}{oqbar}
        \fmfforce{(w,.30h)}{oq}
        \fmfforce{(w,.10h)}{os}
        \fmfforce{(.8w,.40h)}{vg2}
        \fmf{fermion}{ib,vg3,veff,vg1,ou}
        \fmf{phantom}{os,vg4,is}
        \fmffreeze
        \fmf{fermion}{oqbar,vg2,oq}
        \fmf{phantom}{veff,vg1}
        \fmf{dashes}{veff,ow}
        \fmf{gluon}{vg1,vg2}
        \fmf{gluon,tension=0}{vg3,vg4}
        \fmf{fermion,tension=0}{os,vg4,is}
        \fmfv{d.shape=square,d.size=.05h}{veff}
        \fmfv{label=$b$}{ib}
        \fmfv{label=$\bar{u}$}{is}
        \fmfv{label=$u$}{ou}
        \fmfv{label=$\bar{d}$}{oqbar}
        \fmfv{label=$d$}{oq}
        \fmfv{label=$\bar{u}$}{os}
        \fmfv{label=$q$,l.angle=0}{ow}
    \end{fmfgraph*}}
    }
    \end{fmffile}
\end{center}
\caption{\label{fig1} Sketch of QCD factorization in
$B \to \pi\pi\ell\nu$ decays at large dipion masses: (a-b) Leading contributions
from hard gluon exchange; (c) Sample diagram for hard-collinear spectator scattering
corrections.}
\end{figure*}
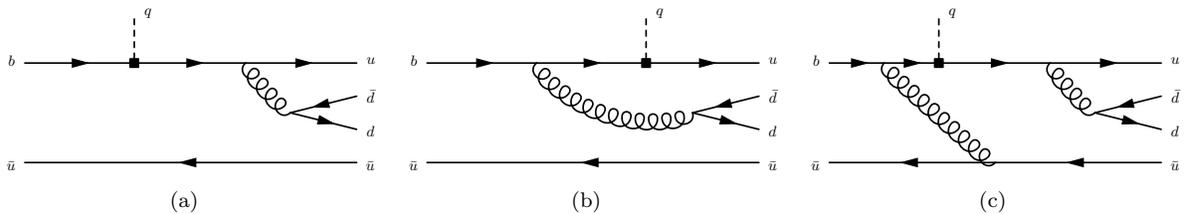
Let us consider the kinematic regime where -- in the $B$-meson rest frame --
the two pions in the hadronic final
state move almost back-to-back, each with large energy, such that
their invariant mass is large, $k^2 \sim {\cal O}(m_b^2)$; see \reffig{phase-space} for an illustration\footnote{
It is to be noted that the QCDF approach for two back-to-back pions also makes use
of SCET techniques for the resummation of large logarithms in higher-order
perturbation theory. In the same way, radiative corrections to form-factor symmetry
relations in SCET can be calculated within the QCDF approach.
}.
In this case, we face a similar situation
as in non-leptonic $B \to \pi\pi$ decays, and thus expect that the
QCD factorization approach from \cite{Beneke:1999br,Beneke:2001ev}
should also be applicable.

Note that in non-leptonic decays, the short-distance quark
transitions are dominantly described by four-quark operators that can
directly induce the leading partonic Fock states required
for $B \to \pi\pi$ transitions in the ``naive'' factorization approach.
For non-leptonic decays, the radiative effects from additional
gluons with virtualities of ${\cal O}(m_b^2)$ (hard) or ${\cal O}(\Lambda_{\rm had} m_b)$
(hard-collinear) thus provide \emph{corrections} to naive factorization.

In $B \to\pi\pi\ell\nu$ the situation is different, as the first non-vanishing
contribution already requires the exchange of a hard gluon in order to produce
the additional $q\bar q$ pair in the final state, see Fig.~\ref{fig1}.
This situation corresponds to the perturbative limit of the 2$\pi$LCDAs
discussed in \cite{Diehl:1999ek} which can then be expressed in terms
of the conventional pion LCDAs.

We thus expect the QCD factorization formula for the dipion form factors
in the considered kinematic limit (and for $m_b \gg \Lambda_{\rm had}$)
to take an analogous form as for non-leptonic $B \to \pi\pi$ decays.
Here at leading term all dipion form factors
would be expressed in terms of a universal $B \to \pi$
form factor, the first inverse moment of the pion LCDA, and simple
kinematic factors. The measurement of the dipion form factors would
thus provide an independent test of the QCD factorization approach,
respectively an independent determination of the relevant hadronic
input parameters.
Radiative corrections from hard and hard-collinear gluon exchange
could be calculated perturbatively,
see Fig.~\ref{fig1}. More details will be provided in \cite{Feldmann:2013prep}.

\subsection{Resonance Contributions}
\label{sec:limits:on-resonance}

Formally, a resonance contribution to $B\to \pi \pi$ form factors can be obtained
using hadronic dispersion relations in the variable $k^2$,
\begin{align}
    \nonumber
    \langle \pi\pi | J_{V-A}^\mu |\bar{B}\rangle
& = \frac{1}{\pi} \int_{4 M_\pi^2}^\infty \dd s\, \frac{\operatorname{Im\,} \langle \pi\pi |J_{V-A}^\mu|\bar{B}\rangle}{s - k^2 - i \eps}\\
    & + \text{subtractions}\,,
\label{eq:dispersion-relation-k2}
\end{align}
with the current $J_{V-A}^\mu = \bar{u}\gamma^\mu(1 - \gamma_5)b$. Insertion of all possible
intermediate states yields a unitarity relation
\begin{multline}
    \label{eq:unitarity-relation-full}
    2 \operatorname{Im\,}\langle \pi\pi |J_{V-A}^\mu |\bar{B}\rangle\\
    = \sum_{H} \int \dd\tau_H \langle \pi\pi|H\rangle\langle H|J_{V-A}^\mu|\bar{B}\rangle\,,
\end{multline}
with integration over the phase space $\tau_H$ and summation over the helicity states of the intermediate
hadronic state $H$.
We single out in this relation $H=R$, with
a resonant one-particle intermediate state $R$, so that the right-hand side contains
the strong coupling $\langle \pi\pi|R\rangle$ of $R$ with two pions, multiplied by the form factors for
$B\to R$ transitions.\\

At this point we must carefully identify
the resonances that emerge in the $k^2$ spectrum, according to the isospin quantum numbers of the
dipion.
In the decay $B^-\to \pi^+\pi^-\ell^-\bar\nu_\ell$ the dipion system is a superposition of the isoscalar $I^G = 0^+$ and
isovector $(I^G,I_3) = (1^+, 0)$ states. In the analogous decay $\bar{B}^0\to \pi^+\pi^0\ell^-\bar\nu_\ell$ and
$B^-\to \pi^0\pi^0\ell^-\bar\nu_\ell$, however,
the pions are purely in the isovector $(I^G,I_3) = (1^+, +1)$ and isoscalar state, respectively.
Altogether, the three hadronic matrix elements for $B\to \pi \pi$ are expressed
in terms of two independent isospin amplitudes. From this we obtain in the isospin symmetry limit the relation
\begin{multline}
\langle \pi^+\pi^- |J_{V-A}^\mu |B^-\rangle
+ \frac{1}{\sqrt{2}} \langle \pi^+\pi^0 |J_{V-A}^\mu |\bar{B}^0\rangle\\
= \langle \pi^0\pi^0 |J_{V-A}^\mu |B^-\rangle\,.
\end{multline}
We consider only resonant contributions due to
the isovector vector mesons $\rho(n)$, as well as the isoscalar scalar mesons $f_0(n)$,
where $n$ denotes the quantum number of radial excitation. We sketch the region of phase space
where the $\rho(n)$ dominate in \reffig{phase-space}. Since we consider only dipion states
up to angular momentum one, we discard resonances with spin larger than one.
Hereafter, we will proceed with the more general case of $B^-\to \pi^+\pi^-\ell^-\bar\nu_\ell$.
The $B^0$ decay can be recovered by omitting the $f_0$ contributions
and adding a relevant isospin factor.

\begin{widetext}
Continuing with \refeq{unitarity-relation-full}, we obtain for the contribution of the $\rho$ intermediate states
\begin{equation}
\label{eq:unitarity-relation-vector}
\begin{aligned}
    \operatorname{Im\,} \langle \pi\pi |J_{V-A}^\mu |\bar{B}\rangle & = -\pi g_{\rho\pi\pi} \delta(M_{\rho}^2 - s) \sum_{a=0,+,-} \left[\bar{k} \cdot \eta(a)\right] \langle \rho(k, \eta(a))|J_{V-A}^\mu|\bar{B}(p)\rangle\,,
\end{aligned}
\end{equation}
with $\eta$ being the polarization vector for the vector state associated with the four-momentum $k$. In the $B$-meson rest frame ($B$-RF)
\begin{equation}
    \eta(\pm)^\mu\Big|_\text{$B$-RF} = \eps(\mp)^\mu\Big|_\text{$B$-RF},\quad
    \eta(0)^\mu\Big|_\text{$B$-RF} = (|\vec{q}\,|, 0, 0, M_B - q_0) / M_V\,,
\end{equation}
see \refapp{kin} for details.
For the $f_0$ state we obtain
\begin{equation}
\label{eq:unitarity-relation-scalar}
    \operatorname{Im\, }\langle \pi\pi |J_{V-A}^\mu |\bar{B}\rangle = \pi g_{f_0\pi\pi} \delta(M_{f_0}^2 - s) M_{f_0}\langle f_0(k)|J_{V-A}^\mu|\bar{B}(p)\rangle\,.
\end{equation}
For both $\rho$ and $f_0$, the above formulae still employ the narrow-width approximation.
The strong couplings are fixed via
\begin{equation}
    \langle \pi\pi|f_0\rangle = g_{f_0\pi\pi} M_{f_0}\,,\qquad
    \langle \pi\pi|\rho(a) \rangle = -(\bar{k} \cdot \eta(a)) g_{\rho\pi\pi}\,,
\end{equation}
for the $f_0$, and for the $\rho$ helicity states $a=\pm, 0$. Note that $g_{\rho\pi\pi} = g_{\rho^0\pi^+\pi^-} =  -g_{\rho^+\pi^+\pi^0}$
due to isospin.\\
\end{widetext}

We use the helicity decomposition of $B\to R$, $R=S,V$ form factors as in \cite{Bharucha:2010im}, adjusted to our
notation and phase convention. By $S(k)$ and $V(k,\eta)$ we shall
denote a hadronic scalar and vector state with momentum $k$ and polarization vector $\eta$, respectively. We define
for the vector resonances
\begin{multline}
\label{eq:formfactor-B-to-V-vector}
\kappa \frac{\sqrt{q^2}}{\sqrt{\lambda_V}} \langle V(k,\eta(\pm)) | \bar{u} \gamma^\mu b | \bar{B}(p)\rangle\\
    = \pm F^{B\to V}_\perp(q^2) \eps^\mu(\pm)\,,
\end{multline}
as well as
\begin{multline}
\label{eq:formfactor-B-to-V-axialvector-long}
    -\kappa \frac{\sqrt{q^2}}{\sqrt{\lambda_V}} \langle V(k,\eta(0)) | \bar{u} \gamma^\mu \gamma_5 b | \bar{B}(p)\rangle\\
    = F^{B\to V}_t(q^2) \eps^\mu(t) - F^{B\to V}_0(q^2) \eps^\mu(0)\,,
\end{multline}
\begin{multline}
\label{eq:formfactor-B-to-V-axialvector-trans}
    -\kappa \frac{\sqrt{q^2}}{\sqrt{\lambda_V}} \langle V(k,\eta(\pm)) |\bar{u}\gamma^\mu \gamma_5 b | \bar{B}(p)\rangle\\
    = F^{B\to V}_\para(q^2) \eps^\mu(\pm)\,,
\end{multline}
and for the scalar resonances
\begin{multline}
\label{eq:formfactor-B-to-S}
    -\frac{\sqrt{q^2}}{\sqrt{\lambda_S}} \langle S(k) | \bar{u} \gamma^\mu \gamma_5 b | \bar{B}(p)\rangle\\
    = F^{B\to S}_t(q^2) \eps^\mu(t) - F^{B\to S}_0(q^2) \eps^\mu(0)\,,
\end{multline}
where we abbreviate $\lambda_R \equiv \lambda(M_B^2, M_R^2, q^2)$ and use an isospin factor
$\kappa = \sqrt{2}$ for $B^-\to \rho^0$ transitions, and $\kappa = 1$ for $\bar{B}^0 \to \rho^+$ transitions.

We express the resonant pole contributions to the $B\to \pi \pi$ form factors in terms of the
$B\to V$ and $B\to S$ form factors. In this way we obtain for all final state polarizations the $P$-wave
contributions
\begin{multline}
    \frac{\sqrt{3}}{\sqrt{2}} \Res F^{(P)}_{\para,\perp}(q^2, k^2)\Big|_{k^2 = P_V}\\
    = \frac{g_{V\pi\pi}}{\kappa} \frac{\sqrt{\lambda_V} M_V}{\sqrt{q^2}} F_{\para,\perp}^{B\to V}(q^2)
\end{multline}
\begin{multline}
    \sqrt{3} \Res F^{(P)}_{0,t}(q^2, k^2)\Big|_{k^2 = P_V}\\
        = \frac{g_{V\pi\pi} \beta_\pi}{\kappa} \frac{\sqrt{\lambda_V} M_V}{\sqrt{q^2}} F_{0,t}^{B\to V}(q^2)\,.
\end{multline}
For the $S$-wave contributions we find
\begin{equation}
    \Res F^{(S)}_{0,t}(q^2, k^2)\Big|_{k^2 = P_S} = g_{S\pi\pi} \frac{\sqrt{\lambda_S} M_S}{\sqrt{q^2}} F_{0,t}^{B\to S}(q^2)\,.
\end{equation}
The total decay width $\Gamma_R$ was added to the pole $P_R = M_R^2 - i M_R \Gamma_R$,
thus yielding standard Breit-Wigner factors
\begin{equation}
    BW_R(k^2) = \frac{1}{[M_R^2 - k^2 - i M_R \Gamma_R]}
\end{equation}
which govern the resonance behavior in the variable $k^2$ close to $k^2 = M_R^2$, $R=\rho(n),f_0(n)$. Note that the widths can be
interpreted as contribution of multihadron states to the imaginary part of $\langle \pi\pi|\bar{B}\rangle$ in $k^2$.
For more details we refer to \cite{Bruch:2004py} where the origin of the $\rho$
width in the pion form factor was discussed in detail.\\

\section{Decay Rate and Angular Analysis}
\label{sec:decay:obs}

In terms of the vector and axial-vector form factors, 
the amplitude for $B \to \pi\pi\ell\nu$ in the SM can be
expressed as
\begin{widetext}
\begin{equation}
    i\calM = i\frac{G_{\rm F} \,  V_{ub}}{\sqrt{2}} 
    \bigg[
    F_0 \, \eps^\mu(0) 
    + \frac{F_\para + F_\perp}{\sqrt{2}} \, \beta_\pi \, \sin\theta_\pi \, e^{+i \phi} \, \eps^\mu(+) 
    + \frac{F_\para - F_\perp}{\sqrt{2}} \, \beta_\pi \, \sin\theta_\pi \, e^{-i\phi} \, \eps^\mu(-)\bigg]
    \big[\bar{u}_\ell \gamma_\mu (1 - \gamma_5) v_\nu\big]\,,
\end{equation}
\end{widetext}
where the helicity form factor for time-like polarization $F_t$ does not contribute in the limit of
massless leptons.
In the following, we find it convenient to express our result in terms of normalized partial-wave amplitudes,
defined from the corresponding partial-wave expansion of the form factors,
\begin{equation}
    A^{(k)}_{n}     = N \, F^{(k)}_n \quad \mbox{(with $n=0,\para,\perp$)}\,,
    \label{eq:def-trans-amp}
\end{equation}
where the normalization factor absorbs kinematic and coupling parameters,
\begin{align}
    N = G_{\rm F} |V_{ub}| \frac{\sqrt{q^2 \beta_\ell \beta_\pi \sqrt{\lambda}}}{\sqrt{3 \cdot 2^{10} \pi^5 M_B^3}}\,,
    \label{eq:def-phase-space-N}
\end{align}
and we restrict our analysis to $k = S,P$ waves in the following.

The five-fold differential decay width for  $\bar{B}\to \pi^+ \pi^0 \ell^- \bar\nu$
then takes a similar form as for the rare FCNC decay $\bar{B}\to \bar{K}\pi \ell^+\ell^-$,
which has received a lot of attention recently \cite{Lu:2011jm,Becirevic:2012dp,Blake:2012mb,Matias:2012qz,Bobeth:2012vn}.
Choosing  $q^2, k^2, \cos\theta_\pi, \cos\theta_\ell$ and $\phi$ as the five independent
kinematic variables,
we obtain
\begin{equation}
\label{eq:decay:obs:ang-dist}
    \frac{8 \pi}{3}\frac{\dd^5\Gamma}{\dd q^2 \,\dd k^2 \,\dd \cos\theta_\pi \,\dd\cos\theta_\ell \,\dd \phi} \equiv J \equiv \sum_{n} J_n \, f_n\,,
\end{equation}
where $J(q^2, k^2, \cos\theta_\pi, \cos\theta_\ell, \phi)$ is decomposed into the angular functions $f_n \equiv f_n(\cos\theta_\pi, \cos\theta_\ell, \phi)$ 
and angular observables $J_n\equiv J_n(q^2, k^2)$.
This notation has been introduced in \cite{Kruger:2005ep} for $B\to K^*(\to K \pi)\ell^+\ell^-$
decays, originally restricted to pure $P$-wave contributions and not taking into account scalar or pseudoscalar
operators (which could be relevant in certain NP models). 
The general case, including a general basis of $b \to u$ operators and interference effects between $S$- and $P$-wave contributions,
can be worked out following  \cite{Altmannshofer:2008dz,Bobeth:2012vn} and reads
\begin{align}
    J & = \big(J_{1s} \sin^2\theta_\pi + J_{1c} \cos^2\theta_\pi + J_{1sc} \cos\theta_\pi\big) \nonumber\\
      & + \big(J_{2s} \sin^2\theta_\pi + J_{2c} \cos^2\theta_\pi + J_{2sc} \cos\theta_\pi\big) \cos 2\theta_\ell \nonumber\\
      & + J_3 \sin^2\theta_\pi \sin^2 \theta_\ell \cos 2\phi \nonumber\\
      & + \big(J_4 \sin 2\theta_\pi + J_{4i} \sin \theta_\pi\big) \sin 2\theta_\ell \cos\phi \nonumber\\
      & + \big(J_5 \sin 2\theta_\pi + J_{5i} \sin \theta_\pi\big) \sin\theta_\ell \cos\phi \cr 
      & + \big(J_{6s} \sin^2\theta_\pi + J_{6c} \cos^2\theta_\pi\big) \cos\theta_\ell \nonumber\\
      & + \big(J_7 \sin 2\theta_\pi + J_{7i} \sin \theta_\pi\big) \sin\theta_\ell \sin\phi \nonumber\\
      & + \big(J_8 \sin 2\theta_\pi + J_{8i} \sin \theta_\pi\big) \sin 2\theta_\ell \sin\phi \nonumber\\
      & + J_{9} \sin^2\theta_\pi \sin^2\theta_\ell \sin 2\phi\,.
\end{align}
Comparing with \refeq{decay:obs:ang-dist} in the SM, we obtain
\begin{align}
    \frac{4}{3} \, J_{1s}  & = \frac{3}{4}  \, \beta_\pi^2 \left(|A^{(P)}_\perp|^2 + |A^{(P)}_\para|^2\right) + \frac{1}{3} \, |A^{(S)}_0|^2\,,\cr
    \frac{4}{3} \, J_{1c}  & = |A^{(P)}_0|^2 + \frac{1}{3}\, |A^{(S)}_0|^2 = -\frac{4}{3}J_{2c}\,,\cr
    \frac{4}{3} \, J_{1sc} & = \frac{2}{\sqrt{3}} \, \Re{A^{(P)}_0 A^{(S)*}_0} = -\frac{4}{3}J_{2sc}\,,\cr
    \frac{4}{3} \, J_{2s}  & = \frac{1}{4} \, \beta_\pi^2 \left(|A^{(P)}_\perp|^2 + |A^{(P)}_\para|^2\right) - \frac{1}{3} |A^{(S)}_0|^2\,,\cr
    \frac{4}{3} \, J_3     & = \frac{1}{2} \, \beta_\pi^2 \left(|A^{(P)}_\perp|^2 - |A^{(P)}_\para|^2\right)\,,
    \label{eq:J9a}
\end{align}
and
\begin{align}
    \frac{4}{3} \, J_4     & = \frac{1}{\sqrt{2}} \, \beta_\pi \Re{A^{(P)}_0 A^{(P)*}_\para}\,,\cr
    \frac{4}{3} \, J_{4i}  & = \frac{\sqrt{2}}{\sqrt{3}} \, \beta_\pi \, \Re{A^{(S)}_0 A^{(P)*}_\para}\,,\cr
    \frac{4}{3} \, J_5     & = \sqrt{2} \, \beta_\pi \, \Re{A^{(P)}_0 A^{(P)*}_\perp}\,,\cr
    \frac{4}{3} \, J_{5i}  & = \frac{2\sqrt{2}}{\sqrt{3}}\,  \beta_\pi \, \Re{A^{(S)}_0 A^{(P)*}_\perp}\,, \cr
    \frac{4}{3} \, J_{6s}  & = 2 \beta_\pi^2 \, \Re{A^{(P)}_\para A^{(P)*}_\perp}\,,\cr
    \frac{4}{3} \, J_{6c}  & = 0\,,
    \label{eq:J9b}
\end{align}
and
\begin{align}
    \frac{4}{3} \, J_7     & = \sqrt{2} \, \beta_\pi \, \Im{A^{(P)}_0 A^{(P)*}_\para} \,,\cr
    \frac{4}{3} \, J_{7i}  & = \frac{2\sqrt{2}}{\sqrt{3}}\, \beta_\pi \, \Im{A^{(S)}_0 A^{(P)*}_\para}\,,\cr
    \frac{4}{3} \, J_8     & = \frac{1}{\sqrt{2}} \, \beta_\pi \, \Im{A^{(P)}_0 A^{(P)*}_\perp}\,,\cr
    \frac{4}{3} \, J_{8i}  & = \frac{\sqrt{2}}{\sqrt{3}} \, \beta_\pi \, \Im{A^{(S)}_0 A^{(P)*}_\perp}\,,\cr
    \frac{4}{3} \, J_9     & = \beta_\pi^2\Im{A^{(P)}_\perp A^{(P)*}_\para}\,. 
    \label{eq:J9c}
\end{align}
Our result for the functions $J_i$ takes an analogous form as found for $\bar{B}\to(\bar{K}\pi)_{S,P}\ell^+\ell^-$
decays in e.g.\ \cite{Becirevic:2012dp,Blake:2012mb}.
Note that the relative strong phases of the dipion form factors can be sizeable,
and we thus keep all the angular observables that involve an imaginary part 
in \refeq{J9c}.

\section{Model-Independent Results}
\label{sec:model-indep-results}

The large number of observables $J_n$ in the angular distribution allows infering
certain information from experimental data, searching for physics
beyond the SM, and testing various theoretical approaches to QCD.

\subsection{Null Tests in and of the SM}
The $V$-$A$ nature of the weak interaction in $b\to u$ transitions can be probed in
$B\to \pi\pi\ell^-\bar\nu_\ell$ decays through two independent, experimental set of null tests.

The first set is given by the theory prediction that
\begin{gather}
    \label{eq:limits:nullJ6c}
    J_{6c} = 0\,,\\
    \label{eq:limits:nullJ12}
    J_{1c} + J_{2c} = 0\,,\quad J_{1sc} + J_{2sc} = 0\,,\\
    \label{eq:limits:nullJ1sc}
    J_{1sc} - \frac{J_5 J_{5i} + 4 J_{8} J_{8i}}{J_{1s} + J_{2s} + 2 J_3} = 0\,,\\
    \label{eq:limits:nullJ6s}
    J_{6s} - \frac{8 J_4 J_{5} + 8 J_{7} J_{8}}{4 J_{1c} - J_{1s} + 3 J_{2s}} = 0\,,\\
    \label{eq:limits:nullJ9}
    J_{9} - \frac{2 J_5 J_{7} - 8 J_{4} J_{8}}{4 J_{1c} - J_{1s} + 3 J_{2s}} = 0\,,
\end{gather}
\begin{multline}
    \label{eq:limits:nullAbsP0Ppara}
    {(-4J_{2c} - (J_{1s} - 3J_{2s}))(J_{1s} + J_{2s} - 2J_3)}\\
    - (16 J_4^2 + 4 J_7^2) = 0\,,
\end{multline}
\begin{multline}
    \label{eq:limits:nullAbsP0Pperp}
    {(-4J_{2c} - (J_{1s} - 3J_{2s}))(J_{1s} + J_{2s} + 2J_3)}\\
    - (4 J_5^2 + 16 J_8^2) = 0\,,
\end{multline}
\begin{multline}
    \label{eq:limits:nullAbsS0Ppara}
    {(J_{1s} - 3 J_{2s})(J_{1s} + J_{2s} - 2J_3)}\\
    - (4 J_{4i}^2 + J_{7i}^2) = 0\,,
\end{multline}
\begin{multline}
    \label{eq:limits:nullAbsS0Pperp}
    {(J_{1s} - 3 J_{2s})(J_{1s} + J_{2s} + 2J_3)}\\
    - (J_{5i}^2 + 4 J_{8i}^2) = 0\,,
\end{multline}
\begin{equation}
    \label{eq:limits:nullFromOverallPhase}
    4 J_9 (J_{5i} J_{4i} + J_{8i} J_{7i}) + J_{6s} (4 J_{8i} J_{4i} - J_{7i} J_{5i}) = 0\,,
\end{equation}
in the absence of $D$-wave or higher partial wave contributions\footnote{
We expect sizable contributions when the dipion mass approaches the mass of the
$f_2$-meson or its radial excitations.}.
Any deviation from \refeq{limits:nullJ6c} would indicate BSM physics
of both scalar and tensor nature, compare \cite{Bobeth:2012vn} in the context
of $\bar{B}\to \bar{K}\pi\ell^+\ell^-$. Breaking of \refeq{limits:nullJ12} can
be achieved by less exotic models which introduce $V$+$A$ interactions.
The relations \refeqs{limits:nullAbsP0Ppara}{limits:nullFromOverallPhase} hold
in the absence of contributions from either scalar or tensor operators. The above
relations are similar to those obtained for the decay $\bar{B}\to \bar{K}^*(\to \bar{K}\pi)\ell^+\ell^-$
in \cite{Egede:2010zc}.

The second set of test only holds in the SCET limit. In that limit
\begin{equation}
    \label{eq:limits:nullSCETJ39}
    \begin{aligned}
        J_{3} & = \order{\Lambda_\text{had}/m_b}\,,\\
        J_{9} & = \order{\Lambda_\text{had}/m_b}\,,
    \end{aligned}
\end{equation}
as well as
\begin{gather}
    \label{eq:limits:nullSCETJ512}
    J_{1s} + J_{2s} - J_{6s} = \order{\Lambda_\text{had}/m_b}\,,\\
    \label{eq:limits:nullSCETJ4578}
    \frac{J_7}{2 J_4} - \frac{2 J_8}{J_5} = \order{\Lambda_\text{had}/m_b}\,,\\
    \label{eq:limits:nullSCETJ4578i}
    \frac{J_{7i}}{2 J_{4i}} - \frac{2 J_{8i}}{J_{5i}} = \order{\Lambda_\text{had}/m_b}\,,
\end{gather}
since the form factors fulfill $F_\perp^{(k)} = F_\para^{(k)} + \order{\Lambda_\text{had}/m_b}$
for all partial waves $k$. Breaking of the relations (\ref{eq:limits:nullSCETJ39})-(\ref{eq:limits:nullSCETJ4578i})
in the SCET limit can only be achieved through either a) subleading corrections to the form
factor relation or b) NP effects in $b\to u$ transitions, such as $V$+$A$ transitions.

\subsection{Accessing Form Factor Ratios and Phase Differences}

We write each form factor $F_i^{(l)}$ in polar form,
\begin{equation}
    F_i^{(l)} = r_i^{(l)} e^{i \phi_i^{(l)}}\,,
\end{equation}
using the moduli $r_i^{(l)}$ and phases $\phi_i^{(l)}$. Given the explicit $V$-$A$ nature of $b\to u$ transitions
in the SM, we can access five phase differences
through ratios of angular observables,
\begin{equation}
\begin{aligned}
    \frac{-2 J_9}{J_{6s}}
        & = \tan(\phi_\para^{(P)} - \phi_\perp^{(P)})\,,\\
    \frac{J_{7}}{2 J_{4}}
        & = \tan(\phi_0^{(P)} - \phi_\para^{(P)})\,,\\
    \frac{J_{7i}}{2 J_{4i}}
        & = \tan(\phi_0^{(S)} - \phi_\para^{(P)})\,,\\
    \frac{2 J_{8}}{J_{5}}
        & = \tan(\phi_0^{(P)} - \phi_\perp^{(P)})\,,\\
    \frac{2 J_{8i}}{J_{5i}}
        & = \tan(\phi_0^{(S)} - \phi_\perp^{(P)})\,,\\
\end{aligned}
    \label{eq:constraints-phases}
\end{equation}
where we employ ten independent angular observables. Moreover, we can access four ratios of moduli $r_i^{(l)} / r_j^{(k)}$
\begin{equation}
    \frac{J_{2sc}}{J_{2c}} = \frac{2 \sqrt{3} r_0^{(S)} / r_0^{(P)}}{3 + (r_0^{(S)} / r_0^{(P)})^2} \cos(\phi_0^{(P)} - \phi_0^{(S)})\,,
    \label{eq:constraints-ratios-1}
\end{equation}
and
\begin{equation}
\begin{aligned}
    \frac{J_{1s} + J_{2s} + 2 J_{3}}{J_{1s} + J_{2s} - 2 J_3} & = \Bigg(\frac{r_\perp^{(P)}}{r_\para^{(P)}}\Bigg)^2\,.\\
    \frac{3 \beta_\pi^2 (J_{1s} - 3 J_{2s})}{2 (J_{1s} + J_{2s} - 2 J_3)} & = \Bigg(\frac{r_0^{(S)}}{r_\para^{(P)}}\Bigg)^2\,,\\
    \frac{3 \beta_\pi^2 (J_{1s} - 3 J_{2s})}{2 (J_{1s} + J_{2s} + 2 J_3)} & = \Bigg(\frac{r_0^{(S)}}{r_\perp^{(P)}}\Bigg)^2\,,
\end{aligned}
    \label{eq:constraints-ratios-2}
\end{equation}
using four further independent observables. Overall this amounts to nine constraints on the form factors
that arise from 14 angular observables. Together with $J_{6c}$ (vanishing in the SM),
$J_{1c,1sc}$ (not independent from $J_{2c,2sc}$ in the SM), and the differential decay width,
\begin{gather}
    \label{eq:constraints-Gamma}
    \frac{\dd \Gamma}{\dd q^2} = J_{1c} - \frac{1}{3} J_{2c} + 2 J_{1s} - \frac{2}{3} J_{2s}\\
    \notag                     \propto \big|V_{ub}\big|^2 \Big[3 (r_\perp^{(P)})^2 + 3 (r_\para^{(P)})^2 + 3 (r_0^{(P)})^2 + (r_0^{(S)})^2 \Big]\,,
\end{gather}
we arrive at 18 angular observables. Thus, the determination of form factor ratios, form factor phases
and the product of form factor moduli and $|V_{ub}|$ as described in \refeqs{constraints-phases}{constraints-Gamma}
exctracts the maximum amount of information from the angular distribution.

\section{Conclusion}
\label{sec:conclusion}

In this paper we have considered the semileptonic decay $B \to \pi \pi \ell
\bar{\nu}_\ell$ in the Standard Model (SM) and analyzed the complete set
of angular observables describing the four-body final state.
Detailed quantitative predictions for
these observables require
genuinely non-perturbative information,
which is encoded in hadronic $B \to \pi\pi$ form factors.
In turn, as we have explored, a full-fledged angular analysis of the decay
will allow one to extract form factor ratios and relative strong phases
from experimental data.
We have also shown that in the soft or collinear limit, the number of independent form
factors is reduced due to heavy quark symmetries in HQET or SCET,
respectively.

The tension in the determination of $|V_{ub}|$ has lead to speculations about
possible non-standard contributions in $b \to u$ transitions.
As we have discussed in this paper, the chiral structure of weak interactions
can be used to identify null tests of the SM in $B\to\pi\pi\ell\bar\nu_\ell$
decay observables;
i.e.\ any violation of the $V$-$A$ structure in $b\to u$ transitions
will show up in modifications of \refeq{limits:nullJ6c} and \refeq{limits:nullJ12}.

Our observations can also be useful for interpolation between different
corners of phase space, where
the resonance structure of the $\pi\pi$-system is described
by phenomenological models, or theoretical calculations based on QCD
factorization, heavy-hadron chiral perturbation theory,
or QCD sum rules are applicable.
Detailed analyses of this kind go beyond the scope of the present paper
and are left for future work.

\section*{Acknowledgments}

This work is supported in parts by the Bundesministerium f\"ur Bildung und Forschung (BMBF),
and by the Deutsche Forschungsgemeinschaft (DFG) within Research Unit FOR 1873
(``Quark Flavour and Effective Field Theories'').

\appendix

\section{Details on the Kinematics}
\label{app:kin}

This appendix shall elaborate on the definitions of kinematic variables in the course
of our calculations, starting with general remarks.

First, we choose the $z$ axis along the flight direction of the dipion system, and consequently
the dilepton system moves along the negative $z$ axis. We also put the dilepton system
into the $x$--$z$ plane.

Second, we make use of a set of virtual polarization vectors $\eps(n)^\mu$, $n = t,\pm,0$,
that fulfill the completeness relations
\begin{equation}
\begin{aligned}
    \eps(n) \cdot q & = 0\qquad n=\pm, 0\\
    \eps(n) \cdot \eps^\dagger(n') & = g_{nn'}\\
    \eps(n)_\mu \eps^\dagger(n')_\nu g_{nn'} & = g_{\mu\nu}\,.
\end{aligned}
\end{equation}
where $g_{nn'} = \text{diag}(+1, -1, -1, -1)$ for $n,n'=t,+,-,0$.

In the following we will discuss the explicit expressions for the various momenta and
polarization vectors in the three frames that are relevant to the decay analysis.

\subsection{The Dilepton Rest Frame}

We describe the dilepton system through its invariant mass $q^2$ as well as the lepton
helicity angle $\theta_\ell$, \ie, the angle between the $\ell^-$ direction of flight and the
the $z$ axis in the dilepton rest frame. We choose the $x$-$z$ plane as the decay
plane of the dilepton system. Thus, we write in the $\ell\nu$ rest frame ($\ell\nu$-RF)
\begin{align}
    q_{1,2}^\mu    \Big|_\text{$\ell\nu$-RF} & = \frac{\sqrt{q^2}}{2} (1, \mp\sin\theta_\ell, 0, \mp\cos\theta_\ell)\,,
\end{align}
and correspondingly
\begin{equation}
\begin{aligned}
    q^\mu      \Big|_\text{$\ell\nu$-RF} & = \sqrt{q^2} (1, 0, 0, 0)\,,\\
    \bar{q}^\mu\Big|_\text{$\ell\nu$-RF} & = -\sqrt{q^2} (0, \sin\theta_\ell, 0, \cos\theta_\ell)\,.
\end{aligned}
\end{equation}

The polarization vectors $\eps^\mu(n)$ take the explicit form
\begin{equation}
\begin{aligned}
    \eps^\mu(t)  \Big|_\text{$\ell\nu$-RF} & = (1, 0, 0, 0)\,,\\
    \eps^\mu(\pm)\Big|_\text{$\ell\nu$-RF} & = (0, 1, \mp i, 0) / \sqrt{2}\,,\\
    \eps^\mu(0)  \Big|_\text{$\ell\nu$-RF} & = (0, 0, 0, -1)\,.
\end{aligned}
\label{eq:polvec-dilepton-rf}
\end{equation}
Comments are due on the choice
of the polarization vectors, especially the signs of $\eps^z(0)$ as well as $\eps^y(\pm)$.
These haven been adopted to obtain longitudinal and right-handed/left-handed polarization
of the $\ell\nu$ system, which moves along the \emph{negative} $z$-axis.\\

\subsection{The $B$-Meson Rest Frame}

In the rest frame of the $\bar{B}$ meson ($B$-RF) we write explicitly
\begin{equation}
\begin{aligned}
    p^\mu\Big|_\text{$B$-RF} & = (M_B, 0, 0, 0)\,,\\
    q^\mu\Big|_\text{$B$-RF} & = (q^0, 0, 0, -|\vec{q}\,|)\,,\\
    k^\mu\Big|_\text{$B$-RF} & = (M_B - q^0, 0, 0, +|\vec{q}\,|)\,.
\end{aligned}
\end{equation}
Since we chose to describe the decay through the invariants $q^2$ and $k^2$, we use
\begin{align}
    q^0        \Big|_\text{$B$-RF} & = \frac{M_B^2 - k^2 + q^2}{2 M_B}\,, &
    |\vec{q}\,|\Big|_\text{$B$-RF} & = \frac{\sqrt{\lambda}}{2 M_B}\,.
\end{align}
Application of a Lorentz boost along $z$-axis from the $\ell\nu$-RF to the $B$-RF
leaves $\eps(\pm)$ invariant, while $\eps(t)$ and $\eps(0)$ are transformed:
\begin{equation}
\begin{aligned}
    \eps^\mu(t)\Big|_\text{$B$-RF} & = (q^0, 0, 0, -|\vec{q}\,|) / \sqrt{q^2}\,,\\
    \eps^\mu(0)\Big|_\text{$B$-RF} & = (|\vec{q}\,|, 0, 0, -q^0) / \sqrt{q^2}\,.
\end{aligned}
\end{equation}

\subsection{The Dipion Rest Frame}

We describe the dipion system through its invariant mass $k^2$ as well as the
pion helicity angle $\theta_\pi$, \ie, the angle between the $\pi^+$ direction of flight
and the $z$ axis in the dipion rest frame ($\pi\pi$-RF). In addition, there is an azimuthal angle
$\phi$ between the dipion and the dilepton decay planes. The planes' normal vectors
are defined in the $B$-RF as $\vec{e}_\pi = (\vec{k}_1 \times \vec{k}_2) / |\vec{k}_1\times \vec{k}_2|$ and
$\vec{e}_\ell = (\vec{q}_2 \times \vec{q}_1) / |\vec{q}_2 \times \vec{q}_1|$, respectively.
Since the angle $\phi$ depends only on the $x$ and $y$ components of $k_1$, $k_2$, $q_1$ and
$q_2$ -- which are invariant under $z$-axis boosts between the $\bar{B}$ rest frame, the dipion rest frame
and the dilepton rest frame -- we find that $\phi$ is the same in all considered
frames of reference laid out in this section. We fix the $x$ axis by requiring
$(q_2)_x > 0$, which implies $\vec{e}_\ell = \vec{e}_y$. From $\phi = 0$ then follows
$\vec{e}_\pi = \vec{e}_y$ and further $(\vec{k}_1)_x < 0$ as well as $(\vec{k}_1)_y = 0$.
The spatial components of $k_1$ therefore point in the negative $x$ direction for $\phi = 0$. Furthermore,
we use $\sin\phi \equiv (\vec{e}_\ell \times \vec{e}_\pi) \cdot \vec{e}_z$ as in \cite{Kruger:2005ep},
from which we infer that $\phi$ is the azimuthal angle of the momentum $k_2$. The
$\pi^+\pi^-$ decay plane is therefore rotated with regard to the dilepton ($x$-$z$)
plane by the angle $-\phi$ around the $z$ axis. From this, one obtains in the dipion rest frame
\begin{align}
    k_1^\mu\Big|_\text{$\pi\pi$-RF}      & = \left(\begin{matrix}E_\pi\\ -\krf \sin\theta_\pi\cos\phi\\ -\krf \sin\theta_\pi\sin\phi\\ +\krf \cos\theta_\pi\end{matrix}\right)\,,\\
    k_2^\mu\Big|_\text{$\pi\pi$-RF}      & = \left(\begin{matrix}E_\pi\\ +\krf \sin\theta_\pi\cos\phi\\ +\krf \sin\theta_\pi\sin\phi\\ -\krf \cos\theta_\pi\end{matrix}\right)\,,\\
\intertext{and consequently}
    k^\mu      \Big|_\text{$\pi\pi$-RF}  & = \left(\begin{matrix}\sqrt{k^2}\\ 0\\ 0\\ 0\end{matrix}\right)\,,\\
    \bar{k}^\mu\Big|_\text{$\pi\pi$-RF}  & = \left(\begin{matrix}0\\ -2\krf\sin\theta_\pi\cos\phi\\ -2\krf\sin\theta_\pi\sin\phi\\ 2\krf \cos\theta_\pi\end{matrix}\right)\,,
\end{align}
with
\begin{align}
    \krf  & \equiv \frac{\beta_\pi}{2} \sqrt{k^2}\,, &
    E_\pi & \equiv \frac{\sqrt{k^2}}{2}\,.
\end{align}
where $\beta_\pi^2 = (k^2 - 4 M_\pi^2)/k^2$.\\

\subsection{Frame-Independent Quantities}

For convenience we present here the scalar products and Levi-Civita contractions that
were used in our calculations, expressed in terms of
the five kinematic variables $q^2$, $k^2$ and the three angles $\theta_\pi$, $\theta_\ell$
and $\phi$. The scalar products read
\begin{align}
    \eps(t) \cdot \bar{q}                     & = 0\\
    \eps(t) \cdot k_{(0)}                     & = \eps(t) \cdot \bar{k}_{(\para)} = 0\\
    \eps(\pm) \cdot \bar{q}                   & = +\frac{\sqrt{q^2}}{\sqrt{2}} \sin\theta_\ell\\
    \eps^\dagger(\pm) \cdot \bar{k}_{(\para)} & = \frac{\beta_\pi \sqrt{k^2}}{\sqrt{2}} \sin \theta_\pi \exp(\pm i \phi)\\
    \eps(0) \cdot \bar{q}                     & = -\sqrt{q^2} \cos\theta_\ell\\
    \eps(0) \cdot k_{(0)}                     & = \frac{\sqrt{\lambda}}{2\sqrt{q^2}}\\
    \eps(0) \cdot \bar{k}_{(\para)}           & = 0\,.
\end{align}
For the contractions with the Levi-Civita we obtain
\begin{align}
    \eps(\eps^\dagger(t),   q, k, \bar{k}) & = \eps(\eps^\dagger(0),   q, k, \bar{k})= 0\\
    \eps(\eps^\dagger(\pm), q, k, \bar{k}) & = \mp i\beta_\pi \frac{\sqrt{\lambda}\sqrt{k^2}}{2\sqrt{2}} \sin\theta_\pi \exp(\pm i \phi)\\
    \eps(q,k,\bar{k},\mu)^2                & = -\frac{\beta_\pi^2}{4} k^2 \lambda \sin^2\theta_\pi
\end{align}
where we abreviate
\begin{equation}
    \eps(a, b, c, d) \equiv a^\mu b^\nu c^\rho d^\sigma \eps_{\mu\nu\rho\sigma},
\end{equation}
and use $\eps^{0123} = -\eps_{0123} = +1$.

\bibliographystyle{apsrev4-1}
\bibliography{references.bib}

\end{document}